%
%
\input amstex
\documentstyle{fic}
\NoBlackBoxes

\topmatter
\title Numerical Quantum Gravity by Dynamical Triangulation \endtitle
\author Kristin Schleich and Donald Witt\\
\address Department of Physics\\ University of British Columbia \\
Vancouver, BC  V6T 1Z1, Canada \endaddress
\endauthor

\leftheadtext{Kristin Schleich and Donald Witt}
\rightheadtext{Numerical Quantum Gravity by Dynamical Triangulation}

\cvol{00}
\cvolyear{0000}
\cyear{0000}

\abstract 
Recently an alternate technique for numerical
quantum gravity, dynamical triangulation, 
has been developed. In this method, the geometry is varied by
adding and subtracting equilateral simplices from the simplicial complex. 
This method overcomes certain difficulties associated with the traditional 
approach in Regge calculus of varying geometry by varying edge lengths. 
However additional complications are introduced: three of the
four moves in dynamical triangulation can violate the simplicial nature of
the complex. Simulations indicate that the rate of these violations
is significant. Thus additional conditions  must be placed on the 
dynamical triangulation moves to 
ensure that the simplicial complex  and its topology are preserved.
\endabstract

\thanks The first author is supported in part by an NSERC grant.
\endthanks
\endtopmatter

\document

\head 1\enspace Introduction\endhead

Numerical evaluations  of path integral expressions for quantum amplitudes can
address many issues not tractable to analytic or perturbative methods. 
It is thus natural to formulate such methods for the study of quantum gravity
and quantum cosmology. A standard
approach to numerically computing quantum amplitudes for gravity is
Regge calculus. In this method, the topology of a history in the path integral
is given in terms of a simplicial complex. Its geometry is specified
by fixing the edge lengths of all simplices in the complex. The geometry is 
varied by varying the edge lengths.
A sum over all geometries is carried out by evaluating
the contribution of these varied geometries to the path integral. 
This approach has been used by several workers
 with some interesting results 
\cite{(See review by Williams and Tuckey [1992], sect. 5)}. However, the 
requirement that simplices remain nondegenerate under variation of edge length 
and the necessity of recomputing
the curvature after each variation are factors that make this technique
somewhat involved.

An alternate approach to numerical computation of quantum gravity, dynamical 
triangulation,  has
been used recently by several workers. This work has concentrated on
searching for phase transitions and characterizing their behavior; recent 
results have found some evidence for such a phase transition in four dimensions 
\cite{(Agishtein and Migdal [1992], Ambjorn and Jurkiewicz [1992], 
Varsted [1992], Brugmann and Marinari [1993])}. 
In the dynamical triangulation approach, the topology is again given in terms
of a simplicial complex.   However, now all
simplices in the complex are fixed to be equilateral. The geometry is then
varied by changing the number of simplices in the complex. Simplices are thus 
always nondegenerate  and additionally always produce the same contribution to 
the curvature which allegedly allows for more rapid numerical evaluation. 

It is key to the use of a simplicial complex in dynamical triangulation
that any changes
in geometry preserve both the simplicial nature of the complex and the 
topology.\footnote{Indeed it is this requirement that restricts
edge lengths in the Regge approach.}
Surprisingly, as described in detail in this paper,
this is not the case for the standard moves
for three dimensional dynamical triangulation: three of the four moves used to 
change the number of simplices in the complex can
violate its simplicial nature. Therefore restrictions on
these moves must be enforced if the simplicial nature of the complex is to be 
preserved by dynamical triangulation. Fortunately, as detailed below, 
such restrictions can be enforced in 3 dimensions.

\head 2\enspace Problems with Dynamical Triangulation\endhead

A discretization of histories appropriate
for quantum gravity is given by a simplicial complex.\footnote{ For a
more detailed presentation of this introductory material, 
see \cite{Schleich and Witt [1993]}.}
First, a {\it n-simplex} is the convex hull of $(n+1)$ affinely
independent points in ${\Bbb R}^{n+1}$.
A 0-simplex is a point, a 1-simplex is an edge, a 2-simplex
is a triangle, a 3-simplex is a tetrahedra and so on.  A simplex that is
the convex hull of a subset of 
the vertices of an n-simplex is a {\it face} of the n-simplex. Then

\definition{Definition} A simplicial complex $K$ is a topological space $|K|$
and a collection of simplices $K$ such that
\roster
\item $|K|$ is a closed subset of some finite dimensional euclidean space,
\item If $F$ is a face of a simplex in $K$, then $F$ is also contained in $K$,
\item If $B$,$C$ are simplices in $K$, then $B\cup C$ is a face of both $B$
and $C$.
\endroster 
The topological space $|K|$ is the union of all simplices in $K$.
\enddefinition

\noindent Simplicial complexes can describe very general topological spaces; 
however those relevant for the current discussion are
manifolds. First, the {\it star} of vertex $v$  is the complex consisting of all 
simplices containing $v$. The {\it link} of $v$ is the complex consisting of the 
subset of simplices in the star of $v$ that do not contain $v$ itself.
Then a closed combinatorial n-manifold is a simplicial complex
for which the link of every vertex is a combinatorial (n-1)-sphere.
The key element in the definition of simplicial complex is that all
elements are uniquely determined by the vertices they contain; no two
elements in the complex are specified by
the same vertices. Were elements of the complex not
uniquely specified by their vertices, additional information
would be needed to differentiate the elements.
Moreover, it is precisely this feature that allows the geometry of a given 
space to be modeled by filling each
simplicial complex with flat space and fixing all edge lengths.
As the geometry is flat on the interior of the simplices, the curvature
is no longer distributed over the space but is carried on (n-2)-simplices.
In dynamical triangulation, one chooses  all n-simplices 
to have equal edge lengths. An advantage of this choice is that the curvature
at each (n-2)-simplex is determined very simply by counting the number $k$ of
n-simplices containing it:
\hbox{$R=2\pi-{k\delta}$}
where $\delta$ is the deficit angle between adjacent faces of an equilateral
n-simplex.
 For  three
dimensions $\delta= \cos^{-1}({1\over 3})$. The Euclidean action for
a closed 3-manifold then reduces to especially simple form,
$ I = \kappa_3 N_3 - \kappa_1 N_1$ where
 $N_i$ is the number of 
i-simplices in the manifold and $\kappa_i$ are constants depending on 
$\delta$, $G$, $\Lambda$ and
the edge length. Thus changes in the action can be computed directly
from changes in the number of simplices in the complex.

Clearly changing the number
of 3-simplices adjacent to a given edge will change the curvature of the space
and correspondingly the geometry of the combinatorial manifold.
Such changes are carried out by a special set of moves.  
In three dimensions, there are four moves as shown in Figure 1; a subdivision
of one tetrahedra into four tetrahedra,  a flip move in which two tetrahedra
sharing a triangle are taken into three tetrahedra sharing an edge, and their 
inverses \cite{(Godfrey and Gross [1991], Ambjorn and Varsted [1992])}.  Note 
that a necessary condition for performing the inverse flip move is that
the edge is in precisely three tetrahedra. Similarly a necessary condition for 
the inverse subdivision move is that the vertex is in precisely four tetrahedra.
The subdivision changes the number of edges and tetrahedra
in the complex by $\Delta N_1 = 4$, $\Delta N_3 = 3$,
and the flip move by $\Delta N_1 = 1$, $\Delta N_3 = 1$. 
Clearly the inverse moves change the counting by the opposite amount. 

At first glance, it would appear that as these moves preserve the boundary of 
the region, they preserve the simplicial nature of the complex. This is not true; 
three of the four moves can violate the simplicial properties of the 3-manifold. 
This fact is not noted in the literature on three dimensional dynamical 
triangulation. It is useful
to demonstrate how such moves can break the complex in two dimensions, where we
can clearly illustrate this. The flip move in two dimensions is illustrated at 
the top of figure 2. Now consider performing the flip move on the 2-manifold at 
the bottom of figure 2; by the nature of the curvature at the flip location,
the edge to be added into the complex is already present. Thus the simplicial
nature of the complex is broken at this location and  the
standard assignment of curvature in Regge calculus cannot be used on the 
resulting space.

In three dimensions analogous violations occur: The flip move can attempt to add 
an edge  that is already in the simplicial 3-manifold. 
The inverse flip move can similarly attempt to add a triangle that is already
present. Finally the inverse subdivision can attempt to add a tetrahedra
that is already in the complex. Although it is hard to illustrate pictorially, 
examples of all of these violations occur for the smallest combinatorial 
3-sphere  consisting of the surface of a 4-simplex, 
$K=\{ v_0,v_1\ldots ,v_4,e_0,e_1\ldots  ,e_9, t_0,t_1\ldots  ,t_9,
q_0,q_1\ldots q_9\}$
with the higher dimensional simplices containing the following vertices:

\vskip 8pt
{\settabs\+$e_0\supset v_0,v_1$\hskip .28in&$e_5\supset v_1,v_3$\hskip .28in
&$t_0\supset v_2,v_3,v_4$\hskip .28in&$t_5\supset v_0,v_2,v_4$\hskip .28in& \cr
\+$e_0\supset v_0,v_1$&$e_5\supset v_1,v_3$\hskip .28in
&$t_0\supset v_2,v_3,v_4$\hskip .28in&$t_5\supset v_0,v_2,v_4$\hskip .28in
&$q_0\supset v_0,v_1,v_2,v_3$\cr
\+$e_1\supset v_0,v_2$&$e_6\supset v_1,v_4$\hskip .28in
&$t_1\supset v_1,v_3,v_4$\hskip .28in&$t_6\supset v_0,v_2,v_3$\hskip .28in
&$q_1\supset v_0,v_1,v_2,v_4$\cr
\+$e_2\supset v_0,v_3$&$e_7\supset v_2,v_3$\hskip .28in
&$t_2\supset v_1,v_2,v_4$\hskip .28in&$t_7\supset v_0,v_1,v_4$\hskip .28in
&$q_2\supset v_0,v_1,v_3,v_4$\cr
\+$e_3\supset v_0,v_4$&$e_8\supset v_2,v_4$\hskip .28in
&$t_3\supset v_1,v_2,v_3$\hskip .28in&$t_8\supset v_0,v_1,v_3$\hskip .28in
&$q_3\supset v_0,v_2,v_3,v_4$\cr
\+$e_4\supset v_1,v_2$&$e_9\supset v_3,v_4$\hskip .28in
&$t_4\supset v_0,v_3,v_4$\hskip .28in&$t_9\supset v_0,v_1,v_2$\hskip .28in
&$q_4\supset v_1,v_2,v_3,v_4$\cr}
\vskip 8pt
All triangles are in two tetrahedra (this is true for all closed 3-manifolds) 
and  are therefore candidates for a flip. Consider $t_0$; it is in $q_3$ and 
$q_4$ and  the vertices that appear in the new edge will be $v_0,v_1$. However, 
this edge is  already in the complex; it is $e_0$. 
Therefore this move cannot be performed.  Similarly, all edges are contained
in precisely 3-tetrahedra (this is not generic to 3-manifolds, but true for
this special case) and are thus candidates for an inverse flip move. Consider 
$e_5$; it is in $q_0,q_2,q_4$ and thus would add a triangle with vertices 
$v_0,v_2,v_4$. But again this triangle is already there; it is $t_5$. Finally, 
the inverse subdivision move at vertex $v_0$ attempts to replace the four 
tetrahedra $q_0,q_1,q_2,q_3$ with one containing vertices
$v_1,v_2,v_3,v_4$. Again this is already there; it is $q_5$. Therefore, all 
of these moves fail. Note that these violations can occur for more general 
3-manifolds;  it is clear that this 3-sphere can be joined to any more general
space to produce a region with edges, triangles and vertices that have similar 
properties.

These violations of the simplicial nature of the complex are serious. 
Fortunately, they are local. Therefore, the simplicial nature of the complex 
can be preserved so long as one tests for these violations before performing a 
move. Such a test is
easy to design:
Check to see
that all new elements to be added to the complex are not currently present. 
Finally, after performing the move, verify that the manifold topology is 
preserved  by  an  algorithm. 
Such a violation testing code has been developed for three dimensional dynamical 
triangulation by the authors.
This code is based on a data structure that encodes the topology of the 
simplicial
complex in hierarchical form. Moves are only implemented after testing for 
violations
of the simplicial nature of the complex.  Data is kept on the frequency of
attempts to perform a move that violates the simplicial nature of the complex.
This code has been run on small scale simulations. Results of these simulations
indicate that the violation rate of the flip move, defined as the percentage of
successful flips on a random sample of all triangles
is about $60$ percent. The violation rate of
the inverse flip move defined as the percentage of
successful inverse flips on a random sample of all edges that
are contained in precisely three tetrahedra 
is around $40$ percent. These high violation frequencies can be intuitively
understood in terms of the subdivision move; it adds 4 edges to the simplex that
contain precisely 3 tetrahedra, but these edges also contain a vertex that 
contains only 4 tetrahedra. Therefore, these edges cannot be used in an 
inverse flip move. It also adds 6 triangles that cannot be used in
a flip move as the new vertex contains only 4 tetrahedra.
 As 
the subdivision move can be performed at any tetrahedra in the complex, 
one expects a continued violation rate for
arbitrarily sized complexes. Finally, the inverse subdivision move
has a very low violation rate, $0.2$ percent. Clearly the violation rates for 
the flip move and inverse flip move are significant.

\head 3\enspace Conclusions\endhead

Any application of dynamical triangulation
should preserve the topology and simplicial nature of the complex. It has been
demonstrated here
that additional conditions on the neighborhoods of the elements subject to
moves must be enforced to do so. Preliminary simulations
indicate that failing to enforce these conditions leads to a 
 high
violation rate. Thus enforcement is important if 
dynamical
triangulation simulations are to implement a discretized version of three 
dimensional gravity.
Furthermore, these results have implications for the four dimensional
case as
the problems discussed in this paper
will appear in four dimensions. These difficulties will be 
magnified
 by the lack of an easily implementable algorithm for testing for a 
 4-manifold.
Thus it may be difficult to verify the additional 
conditions on the moves needed to preserve the simplicial nature 
of the complex are sufficient.

\refstyle{B}

\head Figure Captions\endhead
\noindent Figure 1.{The dynamical triangulation moves
in three dimensions. The first picture illustrates the subdivision move; the
original tetrahedra is replaced with 4 new tetrahedra sharing a common vertex. 
Observe that the boundary of the subdivision is the same as that of the original 
tetrahedra.  The second illustrates the flip move;  two tetrahedra sharing a 
common triangle with vertices $v_0,v_1,v_2$  are replaced with three tetrahedra 
sharing an edge $v_3,v_4$. Again the boundary of the region is 
preserved.}
\vskip 14pt

\noindent Figure 2. {Two dimensional example of a failed
move. The top picture illustrates the flip move in two dimensions.The bottom
picture illustrates the consequences of this move for a vertex 
$v_3$ of high curvature. Note edges specified by $v_0,v_1,v_2$ are not the 
boundary of a triangle in the complex. Before the move, all edges are uniquely 
specified by their vertices. After the move there are two edges with vertices 
$v_0,v_2$. To distinguish them, the new edge has been curved. Curvature is thus
 no longer carried only at vertices and 
the Regge formulas  for it are no longer valid.}

\enddocument